\documentclass[aps,prd,twocolumn,groupedaddress]{revtex4}

\begin{document}


\title{Hawking-Moss Tunneling with a Dirac-Born-Infeld Action}


\author{Daniel Wohns}
\affiliation{Laboratory for Elementary Particle Physics
\\Cornell University, Ithaca, NY 14853}


\date{\today}

\begin{abstract}
The Hawking-Moss tunneling rate for a field described by the
Dirac-Born-Infeld action is calculated using a stochastic approach.
We find that the effect of the non-trivial kinetic term is to
enhance the tunneling rate, which can be exponentially significant.
This result should be compared to the DBI enhancement found in the
Coleman-de Luccia case.
\end{abstract}


\maketitle


Quantum tunneling in gravity is a well studied subject. The
Coleman-de Luccia instanton \cite{CDL} plays an important role in
cosmology, e.g., in gauge theory phase transitions, in inflation and
recently, in the cosmic landscape. In IIB string theory, where the
extra dimensions are compactified, the motion of D3 branes play a
crucial role in brane inflation. As is well-known, the kinetic term
of a D3 brane is given by the Dirac-Born-Infeld action. Recently,
Brown, Sarangi, Shlaer and Weltman showed that Coleman-de Luccia
tunneling with a DBI action can be significantly enhanced when
compared to that with a canonical kinetic term \cite{BSSW}.
Hawking-Moss tunneling arises in the same contexts as Coleman-de
Luccia tunneling when the potential barrier is broad instead of
narrow.  In many cases of interest the potential is not known. A
natural question to ask is what is the impact of a DBI action on the
Hawking-Moss tunneling rate.  Potential applications include both 
inflation and vacuum selection \cite{BSTX}, \cite{TyeCC}, \cite{PE}.  
During the observable stage of inflation observations disfavor large 
field space velocities for the inflaton in models with a DBI action 
\cite{BSTX}.  For this reason we focus on the case where the field 
space velocity is small throughout the tunneling event.

Hawking and Moss used a Euclidean approach in \cite{HawkingMoss} to
calculate the tunneling rate in de Sitter space from a metastable
state A to over a wide barrier with a maximum at B to the true
minimum of the potential C.  Two constant instanton solutions $
\phi(\tau, \mathbf{x}) = \phi_A$, $ \phi(\tau, \mathbf{x}) = \phi_B$
satisfy the classical equations of motion. Claiming that the
tunneling rate is the exponential of the difference in the Euclidean
action of these two instanton solutions $P_{A \rightarrow C} \sim
\exp (-(S(\phi_A) - S(\phi_B))) $ Hawking and Moss found the
tunneling rate to be
\begin{equation}
\label{HM} P_{A \rightarrow C} \sim \exp \bigg( - \frac{3 M_P ^ 4}{8
} \bigg( \frac{1}{V_A} - \frac{1}{V_B} \bigg) \bigg).
\end{equation}

Several problems have been pointed out with this approach in the 
appendices of \cite{KKLT},\cite{TyeCC}. The instanton solution 
$\phi(\tau) = \phi_B$ does not interpolate between the metastable 
state and the true vacuum.  However since the Euclidean scale 
factor $b(\tau) = H ^ {-1} \sin (2 \pi H \tau)$ vanishes at $\tau = 0$ 
and $\tau = H ^ {-1}$ large modifications can be made to the instanton 
solution $\phi(\tau, \mathbf{x}) = \phi_B$ near the endpoints without 
changing the action significantly although the modified solution will no longer solve
the classical equations of motion.  This problem is particularly
severe when there are several metastable vacuua between the initial
metastable state and the true vacuum.  Naively we could pick any of
the instanton solutions sitting at a local maximum between the
initial metastable state and the true vacuum to calculate the
tunneling rate but generically the result depends on this choice of
a local maximum. Additionally the actions $S(\phi_A)$ and
$S(\phi_B)$ are both infinite since they are constant in all of
Euclidean spacetime. The actions can be made finite by only
considering tunneling in a Hubble-sized patch, but we then lose
information about the rest of the universe.  For these reasons we 
will not use the Euclidean approach of \cite{HawkingMoss}.

We find that the tunneling rate is modified to be
\begin{equation}
\label{DBIHM} P_{A \rightarrow C} \sim \exp \bigg( \frac{3 M_p^4}{8}
\bigg(\frac{1}{\gamma V(\phi)} - \frac{(\gamma + 1) (\gamma - 1)}{2
(V(\phi)) ^2} T(\phi) \bigg)\bigg|_{\phi_A}^{\phi_B} \bigg)
\end{equation}
when the inflaton $\phi$ is described by the DBI action.  Here $H$
is the Hubble parameter given by $H^2(\phi) = \frac{8 \pi}{3M_p ^ 2}
\rho$, $T$ is the warped brane tension, and
\begin{equation}
\gamma = \frac{1}{\sqrt{1-\frac{\dot{\phi}^2}{T(\phi)}}} =
\frac{1}{c_s} \geq 1
\end{equation}
where $c_s$ is the sound speed.  For small brane tensions
$T(\phi)/V(\phi) \ll 1$ the tunneling rate in Eq.(\ref{DBIHM}) is
enhanced over the rate in Eq.(\ref{HM}). In the limit that $\gamma =
1$ Eq. (\ref{DBIHM}) reduces to Eq.(\ref{HM}). Our results are valid
in the limit that the curvature of the potential is small compared
to the Hubble scale, that the energy density is dominated by the
potential for the entire region in field space through which the
tunneling occurs, and that the inflaton has a small field space
velocity so that $\gamma \gtrsim 1$.

Instead of the Euclidean approach we will use the more clearly
physically motivated stochastic approach of \cite{GL},
\cite{Starobinskii2}, \cite{GLM}, \cite{Linde}.  The basic physical
idea is that the quantum fluctuations of the short wavelength
components of the inflaton field act as a random force on the long
wavelength parts. If the inflaton is trapped in a metastable
minimum, these fluctuations can drive the field up to a nearby
maximum.  From the maximum the inflaton has a probability of order
unity to roll classically to a different minimum.

We consider a D3 brane probe moving in a type IIB background
described by the DBI action
\begin{equation}
\label{DBI} S = - \int d ^ 4 x a ^ 3 (t) (T \sqrt{1-\frac{\partial
\phi ^2}{T}} - T + V)
\end{equation}
where $T$ is the warped D3 brane tension, $\phi$ is the canonically
normalized inflaton, and $V$ is the potential.  In this theory
$\phi$ has a speed limit because of the non-minimal form of the
kinetic term in Eq.(\ref{DBI}).

We follow closely the treatment of \cite{GL}.  The essential
physical idea is that if we divide the field $\phi(\mathbf{x})$ into
a long-wavelength part $\Phi$ and a short-wavelength part, the
short-wavelength part acts as a random force on the long-wavelength
part.

The average value of the field $\phi$ over the coordinate volume $b
^ 3$ is given by
\begin{equation}
\phi _ b = \frac{1}{(2 \pi) ^{3/2}} \frac{1}{b ^ 3} \int d ^ 3 x
\mathrm{e} ^ {-|\mathbf{x}| ^ 2 / 2 b ^ 2} \phi ( \mathbf{x} ).
\end{equation}
Assuming the metric is flat, we can use the momentum space expansion
\begin{equation}
\phi (\mathbf{x}) = \int \frac{d ^ 3 k}{(2 \pi) ^ {3/2}} \big{ \{ }
a _{\mathbf{k}} \phi _{\mathbf{k}} e ^{i \mathbf{k} \cdot
\mathbf{x}} + a _{\mathbf{k}} ^ \dag \phi _{\mathbf{k}} ^ * e ^{-i
\mathbf{k} \cdot \mathbf{x}} \big{ \} }
\end{equation}
to obtain
\begin{equation}
\label{momexp} \phi _ b = \int \frac{d ^ 3 k}{(2 \pi) ^ {3/2}} e ^{-
k ^ 2 b ^ 2 / 2} \big{ \{ } a _{\mathbf{k}} \phi _{\mathbf{k}} + a
_{\mathbf{k}} ^ \dag \phi _{\mathbf{k}} ^ * \big{ \} }.
\end{equation}

Here $a _{\mathbf{k}}$ and $a _{\mathbf{k}} ^ \dag$ are the ordinary
creation and annihilation operators satisfying $ [ a _{\mathbf{k}},
a _{\mathbf{q}} ^ \dag ] = \delta ( \mathbf{k} - \mathbf{q} )$ . For
calculational simplicity we replace Eq. ~(\ref{momexp}) with

\begin{equation}
\phi _ b = \int \frac{d ^ 3 k}{(2 \pi) ^ {3/2}} \Theta ( -k + b ^
{-1} ) \big{ \{ } a _{\mathbf{k}} \phi _{\mathbf{k}} + a
_{\mathbf{k}} ^ \dag \phi _{\mathbf{k}} ^ * \big{ \} }.
\end{equation}

We are interested in the macroscopic evolution of the field $\phi$
on a de Sitter space background with metric $ds^2 = a ^ 2 (t) (dx ^
2 + dy ^ 2 + dz ^ 2)$.  A string theory construction of such a
background metric is given in \cite{KKLT}.  The coordinate length
associated with the physical length scale $\gtrsim H ^ {-1}$ of
interest is $ b = \frac{1}{\epsilon a} H ^ {-1}$ where $\epsilon \ll
1$. A more precise restriction on $\epsilon$ will be obtained below.
If we let $\Phi$ denote the value of $\phi$ averaged over this
volume, then
\begin{equation}
\Phi = \int \frac{d ^ 3 k}{(2 \pi) ^ {3/2}} \Theta ( -k + \epsilon a
H ) \big{ \{ } a _{\mathbf{k}} \phi _{\mathbf{k}} + a _{\mathbf{k}}
^ \dag \phi _{\mathbf{k}} ^ * \big{ \} }
\end{equation}
and the rate of change of $\Phi$ is
\begin{eqnarray}
\dot{\Phi} = \int \frac{d ^ 3 k}{(2 \pi) ^ {3/2}} \Theta ( -k +
\epsilon a H ) ( a _{\mathbf{k}} \dot{\phi} _{\mathbf{k}} + a
_{\mathbf{k}} ^ \dag \dot{\phi} _{\mathbf{k}} ^ * ) \nonumber \\
+ \epsilon a H ^ 2 \int \frac{d ^ 3 k}{(2 \pi) ^ {3/2}} \delta ( -k
+ \epsilon a H ) ( a _{\mathbf{k}} \phi _{\mathbf{k}} + a
_{\mathbf{k}} ^ \dag \phi _{\mathbf{k}} ^ * ) \nonumber
\\
\equiv \int \frac{d ^ 3 k}{(2 \pi) ^ {3/2}} \Theta ( -k + \epsilon a
H ) \big{ \{ } a _{\mathbf{k}} \dot{\phi} _{\mathbf{k}} + a
_{\mathbf{k}} ^ \dag \dot{\phi} _{\mathbf{k}} ^ * \big{ \} } + g(t).
\end{eqnarray}
We will see that $g(t)$ plays the role of a random force due to the
short-wavelength parts of $\phi$.

To find the equation of motion of $\phi$ we use the results of
\cite{Mukhanov} to calculate the energy density. Using $p = T -T
\sqrt{1-\frac{\dot{\phi}^2}{T}} - V$ and $\rho = 2 X
\partial _X p - p$ where $X = \frac{1}{2} (\nabla \phi)^2$ we obtain
\begin{equation}
\label{rho} \rho = 2 \gamma X + T / \gamma - T + V.
\end{equation}

From Eq.(\ref{rho}) and the Friedman equations we find that the
equation of motion of $\phi$ on the background is
\begin{equation}
\ddot{\phi} + \frac{3 H}{\gamma^2}  \dot{\phi} + (2-\frac{3}{\gamma
^ 2}) \frac{\nabla ^2}{a ^ 2} \phi  + \frac{1}{\gamma ^ 3} V ^
\prime - \frac{(\gamma + 1)(\gamma - 1)}{2 \gamma ^2} T ^ \prime= 0
\end{equation}
where $ ^\prime$ denotes a partial derivative with respect to
$\phi$.

If $\phi = \phi_0$ is a metastable state then we can write $V(\phi)
= V_0 + \frac{m^2}{2} (\phi - \phi_0)^2 - \frac{\lambda}{4} (\phi -
\phi_0)^4$ in a neighborhood of $\phi = \phi_0$.  Assuming $T(\phi)$
is analytic in a neighborhood of $\phi = \phi_0$ we write $T(\phi) =
T_3 (\alpha_0 + \frac{\alpha_2}{2} (\phi - \phi_0) ^ 2 + \cdots)$.
Using these Taylor expansions we see that the mode $\phi_k$
approximately satisfies the equation
\begin{eqnarray}
\ddot{\phi_k} + \frac{3 H}{\gamma^2} \dot{\phi_k} + (\frac{3}{\gamma
^ 2}-2) \frac{k ^2}{a ^ 2} \phi_k + \frac{1}{\gamma ^ 3} (m^2 \phi_k
- \lambda \langle \phi^2 \rangle \phi_k) \nonumber \\
- \frac{(\gamma + 1)(\gamma - 1)}{2 \gamma ^2} T_3 (\alpha_2 \phi_k
+ \cdots) = 0.
\end{eqnarray}

If the short-wavelength modes of $\phi$ are to act as a stochastic
force on the spatially averaged $\Phi$, then these modes should obey
the equation of a free field.  For modes with $ k \gg \epsilon a H $
only the first three terms are important if three conditions are
satisfied:

1) $m^2 / \gamma ^ 3 \ll (3 / \gamma ^ 2 - 2) k ^ 2 / a ^ 2$

2) $\lambda \langle \phi ^ 2 \rangle / \gamma ^ 3 \ll (3 / \gamma ^
2 - 2) k ^ 2 / a ^ 2$

3) $\frac{(\gamma+1) (\gamma - 1)}{2 \gamma^2} T_3 (\alpha _ 2 +
\cdots) \ll (3 / \gamma ^ 2 - 2) k ^ 2 / a ^ 2$.

The first two conditions restrict the shape of the potential for
which our approximation is valid.  These conditions are weaker than
the conditions for the validity of the Hawking-Moss tunneling rate (\ref{HM})
obtained by \cite{GL}.  Our conditions reduce to the conditions of
\cite{GL} when $\gamma = 1$.  Condition 3 is always valid for small
field space velocities (sufficiently close to $\gamma = 1$) and we
know that $\gamma \gtrsim 1$ in a metastable state. Condition 2 is
always valid in the neighborhood of a metastable state because the
effective square of the mass $M ^ 2 = m^2 - \lambda \langle \phi ^ 2
\rangle$ is always positive. This positivity combined with condition
1 implies
\begin{equation}
\lambda \langle \phi ^ 2 \rangle < m ^ 2 \ll (3 \gamma - 2 \gamma ^
3) k ^ 2 / a ^ 2
\end{equation}
By \cite{Mukhanov} the expectation value of $\phi ^ 2$ is given by
\begin{equation}
\langle \phi ^ 2 \rangle = \frac {3 H ^ 4}{8 \gamma ^ 2 \pi ^ 2 m ^
2}
\end{equation}
which gives us a bound on $\lambda$
\begin{equation}
\lambda < \frac{8 \pi ^ 2}{3} \frac{m ^ 4 \gamma ^ 2}{H ^ 4} \ll
\frac{8 \pi ^ 2}{3} (3 \gamma^3 - 2 \gamma ^ 5 ) \epsilon ^ 4.
\end{equation}

When our three conditions are satisfied the modes with $k \gtrsim
\epsilon a H$ satisfy the equation
\begin{equation}
\label{eom} \ddot{\phi_k} + \frac{3 H}{\gamma^2} \dot{\phi_k} +
(\frac{3}{\gamma ^ 2}-2) \frac{k ^2}{a ^ 2} \phi_k  = 0
\end{equation}
where $a = e ^ {H t}$.  Using the results of \cite{Mukhanov}
\begin{equation}
\phi _ k \approx - \imath \frac{H}{\gamma (2 k ^ 3) ^ {1/2}}
\end{equation}
at the sound horizon.  Substituting the expansion of $\phi$ in
momentum space into our equation of motion we find
\begin{eqnarray}
\label{int} \int \frac{d ^ 3 k}{(2 \pi) ^ {3/2}} & & \nonumber \\
\big{ \{ } a
_{\mathbf{k}} \big{(} \ddot{\phi} _{\mathbf{k}} + \frac{3 H}{\gamma
^ 2} \dot{\phi} _{\mathbf{k}} + \big{(}\frac{3}{\gamma ^ 2}
 - 2\big{)} \phi _{\mathbf{k}} \big{)} e ^{i \mathbf{k} \cdot \mathbf{x}} +
\textrm{H.C.} \big{ \} } & & \nonumber \\
 + \frac{1}{\gamma ^ 3} V ^ \prime - \frac{(\gamma + 1)(\gamma -
1)}{2 \gamma ^2} T ^ \prime = 0 & &
\end{eqnarray}
The integral in Eq.(\ref{int}) is equal to
\begin{eqnarray}
\int \frac{d ^ 3 k}{(2 \pi) ^ {3/2}} \Theta ( \epsilon a H - k )
\big{ \{ } a _{\mathbf{k}} \big{(} \ddot{\phi} _{\mathbf{k}} +
\frac{3 H}{\gamma ^ 2} \dot{\phi} _{\mathbf{k}} + \nonumber \\
\big{(}\frac{3}{\gamma ^ 2}
 - 2\big{)} \phi _{\mathbf{k}} \big{)} e ^{i \mathbf{k} \cdot \mathbf{x}} +
\textrm{H.C.} \big{ \} }
\end{eqnarray}
because
\begin{eqnarray}
\int \frac{d ^ 3 k}{(2 \pi) ^ {3/2}} \Theta (k - \epsilon a H )
\big{ \{ } a _{\mathbf{k}} \big{(} \ddot{\phi} _{\mathbf{k}} +
\frac{3 H}{\gamma ^ 2} \dot{\phi} _{\mathbf{k}} + \nonumber \\
\big{(}\frac{3}{\gamma ^ 2}
 - 2\big{)} \phi _{\mathbf{k}} \big{)} e ^{i \mathbf{k} \cdot \mathbf{x}} +
\textrm{H.C.} \big{ \} }
\end{eqnarray}
vanishes identically by Eq.(\ref{eom}). In Eq. (\ref{int}) the first
derivative term is the most important if $\gamma \gtrsim 1$ which
must be the case if the inflaton starts out in a metastable state.
We assume $\gamma \gtrsim 1$ for the entire region in field space
through which tunneling occurs. Thus
\begin{eqnarray}
\int \frac{d ^ 3 k}{(2 \pi) ^ {3/2}} \Theta ( \epsilon a H - k )
\big{ \{ } a _{\mathbf{k}} \dot{\phi} _{\mathbf{k}} e ^{i \mathbf{k}
\cdot \mathbf{x}} + \textrm{H.C.} \big{ \} } \nonumber \\
= - \frac{1}{3 H}
(\frac{1}{\gamma} V ^ \prime - \frac{(\gamma + 1)(\gamma - 1)}{2} T
^ \prime)
\end{eqnarray}
Averaging this equation over the volume $b ^ 3$ we find that
\begin{equation}
\label{Langevin} \dot{\Phi} = - \frac{1}{3 H}
\bigg{(}\frac{1}{\gamma} \frac{\partial V}{\partial \Phi}-
\frac{(\gamma + 1)(\gamma - 1)}{2} \frac{\partial T}{\partial
\Phi}\bigg{)} + g(t)
\end{equation}
where
\begin{equation}
g(t) = \epsilon a H ^ 2 \int \frac{d ^ 3 k}{(2 \pi) ^ {3/2}} \delta
( -k + \epsilon a H ) \big{ \{ } a _{\mathbf{k}} \phi _{\mathbf{k}}
+ a _{\mathbf{k}} ^ \dag \phi _{\mathbf{k}} ^ * \big{ \} }.
\end{equation}

Equation (\ref{Langevin}) is a Langevin equation with a random force
$g(t)$. The correlation functions $\langle g(t) \rangle$, $\langle
g(t_1) g(t_2) \rangle$, etc. characterize the statistical properties
of $g(t)$. We compute these functions by averaging over the vacuum
state $| \rangle$ that satisfies $a _{\mathbf{k}} | \rangle = 0$.
 Clearly all of the odd correlation functions vanish.  The two-point
 correlation function is given by
 \begin{eqnarray}
\langle g ( t _ 1 ) g ( t _ 2 ) \rangle & = & \epsilon ^ 2 H ^ 4 a _
1 a _ 2 \int \frac{d ^ 3 k d ^ 3 q}{(2 \pi ) ^ 3} 
\delta ( k -
\epsilon a _ 1 H ) \delta ( q - \epsilon a _ 2 H ) \nonumber \\
& &\langle a
_{\mathbf{k}} a _{\mathbf{q}} ^ \dag \rangle \phi _{\mathbf{k}} \phi
^ * _{\mathbf{q}} \nonumber
\\
& = & \frac{\epsilon ^ 2 H ^ 6 a _ 1 a _ 2}{(2 \pi) ^ 2 \gamma ^ 2} \int
\frac{dk}{k} \delta (k - \epsilon a _ 1 H) \delta (k - \epsilon a _
2 H) \nonumber \\
& = & \frac{\epsilon ^ 2 H ^ 6 a _ 1 a _ 2}{(2 \pi) ^ 2 \gamma ^ 2}
\frac{1}{\epsilon a _ 1 H} \delta (\epsilon a_1 H - \epsilon a _ 2
H) \nonumber \\
& = & \frac{H ^ 3}{4 \pi ^ 2 \gamma ^ 2} \delta (t_1 - t _2)
 \end{eqnarray}
It can be shown by induction that $\langle g(t_1) \cdots g(t _ n)
\rangle = \sum \prod \langle g ( t _ i) g ( t _ j) \rangle$ for all
even $n$ where the sum is taken over all possible products of
two-point functions.  Therefore $g(t)$ is a Gaussian variable and
Eq.(\ref{Langevin}) leads to the standard Fokker-Planck equation
\begin{equation}
\label{FokkerPlanck} \frac{\partial \rho}{\partial t} =
\frac{\partial}{\partial \Phi} \bigg( \frac{1}{3H} \bigg( \frac{1}
{\gamma} \frac{\partial V}{\partial \Phi} - \frac{(\gamma + 1)
(\gamma - 1) }{2} \frac{\partial T}{\partial \Phi} \bigg) \rho
\bigg) + D \frac{\partial ^ 2 \rho}{\partial \Phi ^ 2}
\end{equation}
with $D = H ^ 3 / 8 \pi ^ 2 \gamma ^2$.

There is a finite probability for a particle initially in a
metastable state at $\phi = \phi_A$ to stochastically climb up the
potential to a nearby maximum at $\phi = \phi _ B$.  Once at the
maximum the particle has a probability of order unity to classically
roll down to an adjacent minimum $\phi = \phi_C$.  By integrating
the Fokker-Planck equation
\begin{equation}
\frac{\partial \rho}{\partial t} = \frac{\partial}{\partial \Phi} \bigg( \frac{1}{3H} 
\frac{\partial V}{\partial \Phi} \rho
\bigg) + D \frac{\partial ^ 2 \rho}{\partial \Phi ^ 2}
\end{equation}
in \cite{Starobinskii} it was found that
the mean time during which a particle initially at $\phi = \phi_A$
passes over the barrier at $\phi = \phi_B$ of height $\Delta V$ is
given by
\begin{equation}
\Delta t \sim \exp \bigg( \int _ {\phi_A}^{\phi_B} d \phi
\frac{\partial}{\partial \phi} \frac{V(\phi)}{3 H(\phi) D(\phi)}
\bigg)
\end{equation}
up to some subexponential prefactors. Using the same technique and 
(\ref{FokkerPlanck}) we see that
\begin{equation}
\Delta t \sim \exp \bigg( \frac{8 \pi ^ 2}{3}
\bigg(\int_{\phi_A}^{\phi_B} d \phi \frac{\partial}{\partial \phi}
\bigg(\frac{V}{\gamma H ^ 4}
 - \frac{(\gamma + 1) (\gamma -
1)}{2 H ^ 4} T \bigg) \bigg) \bigg)
\end{equation}
if the argument of the exponential is large.  Equivalently the
probability per unit volume is given by
\begin{eqnarray}
\label{prob} P_{A \rightarrow C} &\sim& \exp (-B_{HMDBI}) \nonumber \\
&\sim &\exp \bigg( \frac{3 M_P^4}{8}
\bigg(\frac{1}{\gamma V} - \frac{(\gamma + 1) (\gamma - 1)}{2 V ^2}
T \bigg)\bigg|_{\phi_A}^{\phi_B} \bigg)
\end{eqnarray}
so the effect of the DBI action is to modify the tunneling rate from
the result of \cite{HawkingMoss}
\begin{eqnarray}
\label{HM_prob}
P_{A \rightarrow C} &\sim &\exp (-B_{HM}) \nonumber \\
&\sim &\exp \bigg(- \frac{3M_P ^ 4}{8} \bigg(
\frac{1}{V(\phi _ A)} - \frac{1}{V(\phi _ B)} \bigg) \bigg).
\end{eqnarray}
For example in a warped type IIB compactification with fluxes \cite{KKLT} and D7-branes wrapped on
4-cycles, the potential for a probe D3-brance can have discrete minima in the angular
directions of the compact space \cite{BDKM}.  If at a fixed radial position at the bottom of
the throat the D3-brane is in a false vacuum, it can tunnel in an angular direction.  
If we have $T = (1.0* 10^{-5} M_P)^4$, $V(\phi _ A) = (3.00 * 10^{-4} M_P)^4$, 
$V(\phi _ B) = (3.01 * 10^{-4} M_P)^4$, $\gamma _ A = 1.02$, and $\gamma _ B = 1.01$, then
$B_{HM}/B_{HMDBI} \sim 3.91$ so tunneling is exponentially faster than one would expect
using (\ref{HM_prob}).
This result could have important applications for inflation when the 
inflaton is described by the DBI action, or for vacuum selection in 
the landscape.  We leave these implications to future work.

After completing this work the author became aware of \cite{TW} in
which a result is derived that agrees with Eq.(\ref{DBIHM}) in the
limit that $T/V \ll 1$.  See also \cite{Brown} for a related discussion.

\begin{acknowledgments} I would like to thank D. Balick, C. Csaki,
L. McAllister, and J. Xu for useful and interesting discussions.  I
am especially grateful to S.-H. H. Tye for suggesting the topic and
for providing numerous suggestions on drafts.  This research was
supported in part by the National Science Foundation under Grant
PHY-0355005.
\end{acknowledgments}



\begin{thebibliography}{99}
\bibitem{CDL} S. R. Coleman, ``The Fate Of The False Vacuum. 1. Semi-
classical Theory,¡± Phys. Rev. D 15, 2929 (1977).
\bibitem{BSSW}  A. R. Brown, S. Sarangi, B. Shlaer, and A. Weltman, ``A Wrinkle in Coleman - De Luccia,"
Phys. Rev. Lett. 99, 161601 (2007); arXiv:hep-th/0706.0485.
\bibitem{BSTX} R. Bean, S. E. Shandera, S.-H. H. Tye, J. Xu, ``Comparing Brane Inflation to WMAP", JCAP 2007;
 [arXiv:hep-th/0702107].
\bibitem{TyeCC} S.-H. H. Tye, ``A Renormalization Group Approach to the Cosmological Constant
Problem," [arXiv:hep-th/arXiv:0708.4374].
\bibitem{PE} D. Podolsky and K. Enqvist, ``Eternal inflation and localization on the landscape," [arXiv:0704.0144v3].
\bibitem{HawkingMoss} S. W. Hawking and I. G. Moss, ``Supercooled Phase Transitions
In The Very Early Universe," Phys. Lett. B 110, 35 (1982).
\bibitem{KKLT} S. Kachru, R. Kallosh, A. Linde and S. P. Trivedi, ``De Sitter vacua
in string theory," Phys. Rev. D 68, 046005 (2003),
[arXiv:hep-th/0301240].
\bibitem{GL} A.S. Goncharov and A.D. Linde, ``Tunneling in an
expanding universe: Euclidean and Hamiltonian approaches," Sov. J.
Part. Nucl. \textbf{17} 369 (1987).
\bibitem{Starobinskii2} A. A. Starobinsky,
``Stochastic DeSitter (Inflationary) Stage in the Early Universe,"
in \textit{Field Theory, Quantum Gravity and Strings}, edited by
H.J. De Vega and N. Sanchez (Springer New York 1986).
\bibitem{GLM} A. S. Goncharov, A. D. Linde and V.
F. Mukhanov, ``The Global Structure Of The Inflationary Universe,"
Int. J. Mod. Phys. A 2, 561 (1987).
\bibitem{Linde} A. D. Linde,
Particle Physics and Inflationary Cosmology, (Harwood, Switzerland
1990), [arXiv:hep-th/0503203]
\bibitem{Starobinskii} A. A. Starobinskii, ``Fundamental'nye
vzaimodeistviya," Moscow State Pedagogical Institute, Moscow, 55
(1984).
\bibitem{Mukhanov} V. Mukhanov, ``Physical Foundations of
Cosmology," Cambridge University Press, Cambridge, 2005.
\bibitem{TyeReview} S.-H. H. Tye, ``A new view of the cosmic landscape,"
[arXiv:hep-th/0611148].
\bibitem{SilversteinTong} E. Silverstein and D. Tong, ``Scalar Speed Limits and
Cosmology: Acceleration from D-cceleration," Phys. Rev. D 70, 103505
(2004); [arXiv:hep-th/0310221].
\bibitem{BDKM} D. Baumann, A. Dymarsky, I. R. Klebanov, and L. McAllister, 
``Towards an Explicit Model of D-brane Inflation,"  [arXiv:hep-th/0706.0360].
\bibitem{TW}  A. J. Tolley and M. Wyman, ``Stochastic Inflation Revisited: Non-Slow Roll Statistics and DBI
Inflation," [arXiv:hep-th/0801.1854].
\bibitem{Brown} A. Brown, ``Brane tunneling and virtual brane-antibrane pairs"; [arXiv:0709.3532v1].
\end{thebibliography}
\end{document}